\documentclass{modelica}
\hypersetup{%
  pdftitle  = {Discontinuity-Aware Solvers for Vapor-Compression Cycles},
  pdfauthor = {Shreyas Ekanathan, Avinash Subramanian, Oscar Smith, Yingbo Ma, Christopher Rackauckas, Christopher R. Laughman},
  pdfsubject = {American Modelica and FMI Conference 2026},
  pdfkeywords = {Vapor-compression cycles, HVAC, Differential equations, Dyad},
  colorlinks,
  linkcolor=black,
  urlcolor=black,
  citecolor=black,
  pdfpagelayout = SinglePage,
  pdfcreator = pdflatex,
  pdfproducer = pdflatex}

\usepackage[utf8]{inputenc}
\usepackage{multirow}
\usepackage{graphicx}
\usepackage{tabularx} 
\usepackage{subcaption}
\usepackage{stfloats}
\usepackage{amsmath}
\usepackage{amssymb}

\usepackage{listings}
\usepackage{xcolor}

\lstdefinelanguage{Dyad}{
  morekeywords={component, parameter, variable, relations, end,
                extends, connect, initial, metadata, if, then, else},
  sensitive=true,
  morecomment=[l]{\#},
  morestring=[b]",
}

\makeatletter
\renewcommand\paragraph{\@startsection{paragraph}{4}{\z@}%
  {1.2ex \@plus 0.3ex \@minus 0.2ex}%
  {-0.7em}%
  {\normalfont\normalsize\bfseries}}
\makeatother

\newcommand{\pd}[2]{\frac{\partial #1}{\partial #2}}
\newcommand{\dd}[2]{\frac{d #1}{d #2}}
\newcommand{\mdot}{\dot{m}}

\begin{document}
\thispagestyle{empty}

\title{Discontinuity-Aware Solvers for Vapor-Compression Cycles}
\author[1, 3]{Shreyas Ekanathan}
\author[1]{Avinash Subramanian}
\author[1]{Oscar Smith}
\author[4]{Yingbo Ma}
\author[1]{Christopher Rackauckas}
\author[2*]{Christopher R. Laughman}

\affil[1]{JuliaHub, Inc. {\small\texttt{\{shreyas.ekanathan, avinash.subramanian, oscar.smith, chris.rackauckas\}@juliahub.com}}}
\affil[2]{Mitsubishi Electric Research Laboratories, Cambridge, MA, USA {\small\texttt{laughman@merl.com}}}
\affil[3]{University of Texas at Austin}
\affil[4]{Independent Researcher}

\maketitle\thispagestyle{empty} 
\abstract{
Physics-based models of vapor-compression cycles are often formulated as large, stiff DAE systems with derivative discontinuities introduced at the points of phase change. We use an implementation of a discontinuity-aware ODE and DAE solver scheme that uses domain-specific, user-declared information to compute appropriate time steps for more efficient integration. This scheme is made open-source within \texttt{DifferentialEquations.jl}. We demonstrate how discontinuity information can be specified symbolically in a declarative modeling language such as \texttt{Dyad} and passed to the solver to address the numerical challenges posed. We apply this discontinuity-aware technique to two applications: First, a scalar ODE with two discontinuities, and then a model of a full vapor-compression cycle.  The discontinuity-aware solver yields significantly fewer step rejections and improved solver times, particularly with the \texttt{Rodas5P} solver at loose tolerances, as compared to standard error control, suggesting that exposing physical discontinuity information from equation-oriented models to adaptive solvers can improve practical simulation performance without requiring reassembly of the model.
}

\section{Introduction}

Vapor-compression cycles (VCCs) represent a key technology that serves as a foundation for many heating and cooling systems used in residential, commercial, and industrial buildings today, such as in heat pumps and air-conditioning systems.  These systems play an essential role in maintaining comfort in buildings and regulating industrial thermal processes, and also provide an important capability in the continuing trend of electrification as a means of integrating renewable energy sources into the electrical grid.  Given ongoing climate trends, these systems are becoming more common throughout the world in both existing and emerging markets. 

Considering the wide geographic diversity of equipment installations and the range of applications, model-based design processes for equipment that incorporates vapor-compression cycles is becoming increasingly important to manage their dynamic behavior in the face of changing operating conditions.  Whereas common design processes for these cycles tend to focus on experimental trial-and-error methodologies, system-level modeling and simulation is a valuable tool when used to pursue high system-of-systems efficiencies and reduce the dependence of the design process on time-consuming and expensive experimental laboratory work.  These tools can improve the design process across many different system aspects, including refrigerant selection, component sizing, control system design, and initial design evaluations in the larger system-of-systems context.  

A variety of approaches have been used to describe the spatially and temporally varying behavior of these systems as a set of computational models~\cite{rasmussen2012dynamic1, rasmussen2012dynamic2}.  While the underlying physics of these systems might be most accurately characterized as a set of partial differential algebraic equations (PDAEs), these are often simplified to form either a set of ordinary DAEs or a set of ordinary differential equations (ODEs).  DAEs often arise in these systems due to the algebraic relations between the refrigerant properties, as well as from algebraic equations from flow and energy constraints formed when interconnecting component-based models.  This modeling approach has been popularized by the Modelica language~\cite{MLSv36} and recently by the Dyad language~\cite{rackauckas2025dyad}, in which component models are created using an equation-oriented declarative language, and then interconnected to form larger system-level models.  We employ this approach to construct these VCC models, which are based on well-known previous work~\cite{qiao2015}.

While a substantial body of work exists for the numerical solution of general DAEs~\cite{brenan1995numerical, petzold1982differential}, the numerical solution of DAE models of VCCs pose a number of distinct challenges.  These models are often quite large, due to the discretization of the heat exchangers into a number of control volumes to approximate the behavior of the partial differential equations describing the conservation of mass, momentum, and energy within the refrigerant flow and across the pipe wall.  In addition, these models contain a variety of  nonlinear relationships, such as between the refrigerant properties (e.g., temperature, density) or in the closure relations relating the pressure drop to the refrigerant flow, which must be accommodated when solving these models.

\begin{figure}[t]
\centering
\includegraphics[width=0.48\textwidth]{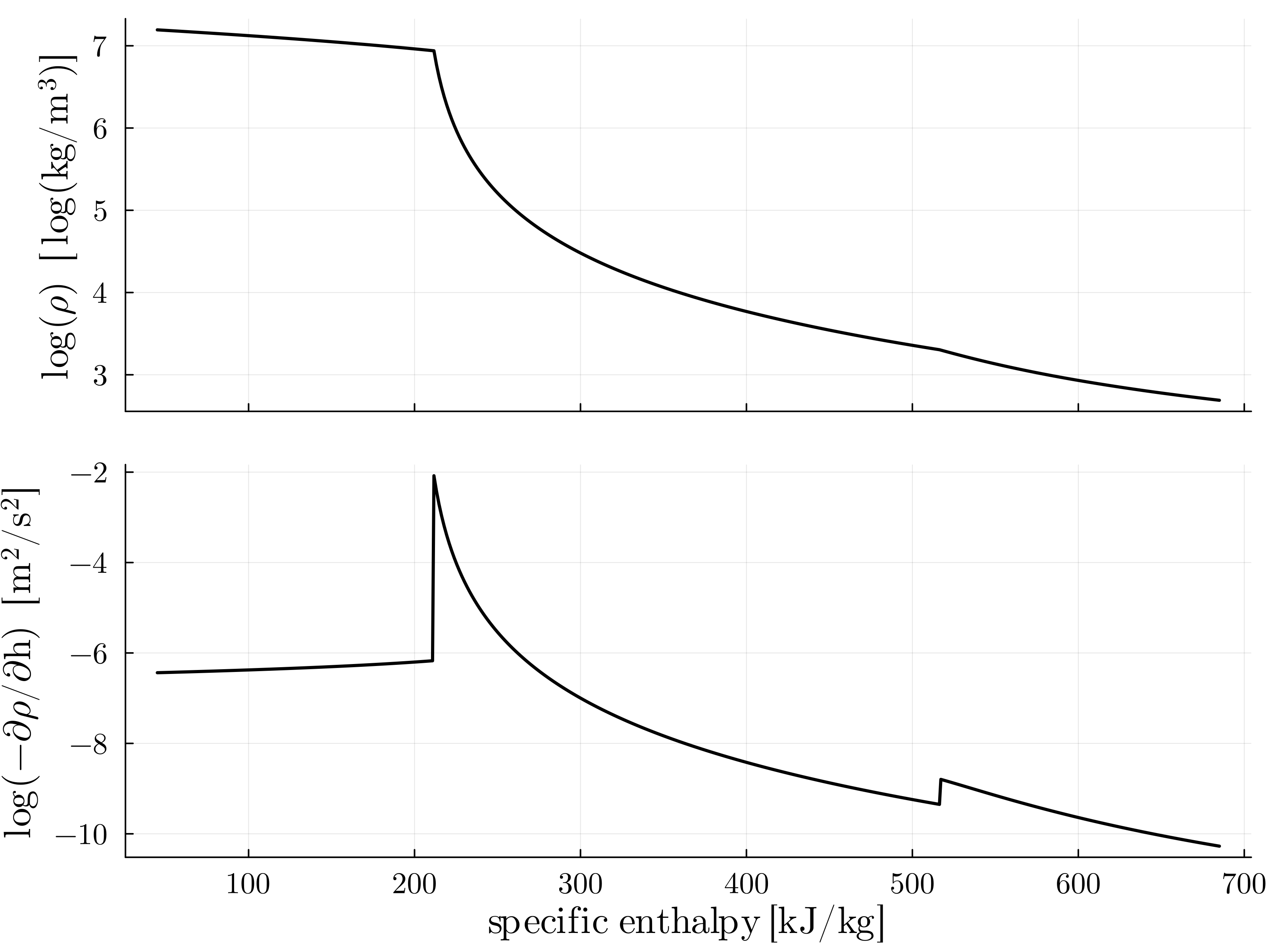}
\caption{Plot of $\log(\rho)$ and $\log(\left.-\partial \rho / \partial h\right|_P)$ for the refrigerant R32 at 1 MPa over a range of specific enthalpies, illustrating the derivative discontinuities at the bubble and dew points.}
\label{fig:density_1MPa}
\end{figure}
 
The $C^0$ behavior of the refrigerant in the two-phase (condensing or evaporating) flows also introduces numerical challenges to the solver. This behavior is observed when integrating the differential equations related to the conservation of mass, which are often formulated in terms of the density (as the cross-sectional area of the tubes in the heat exchangers are constant).  As seen in the example provided in Figure~\ref{fig:density_1MPa}, this density has discontinuous derivatives at the boundaries of the two-phase region, known as the bubble point (between the liquid and two-phase regions) and the dew point (between the two-phase and vapor regions).  Numerical solvers that integrate the differential equations of the model are often designed under the assumption that the underlying model is smooth, and may either become extremely slow when encountering such derivative discontinuities, or even crash or produce erroneous output in these situations. This is because the solver steps past the discontinuity, incurs a large local error, rejects the step, reduces the step size, and repeats this process many times before successfully crossing the trouble spot~\cite{gear1984}.

Several techniques have been developed for solving dynamical systems with discontinuities~\cite{dieci2012}, which arise in two distinct varieties: time-triggered and state-triggered.  Time-triggered discontinuities occur at prescribed times that are known in advance and so are easier to handle, as the integrator can step to the discontinuity location and restart integration ~\cite{mork2025}. State-triggered discontinuities, on the other hand, require the determination of the time location of the discontinuity during integration.  While most numerical integration methods are constructed on the assumption that the solution is smooth, these discontinuities violate this assumption. Gustafsson illustrates the effect of these discontinuities by showing that a naive integrator can spend a lot of computation on repeated step rejections to handle the large error introduced, whereas an informed integrator may directly step to it~\cite{gustafsson1994traps}. 

A number of works have addressed the numerical integration of dynamical systems with state-triggered discontinuities.  Carver~\cite{carver1978efficient} and Cellier~\cite{cellier1979combined} published seminal work on marking the discontinuity using zero-crossing functions, which are checked by the integrator for sign changes that indicate when the discontinuity is crossed. After crossing the discontinuity, the integrator is restarted with the chosen set of equations for the branch.  One distinct mode of failure for this approach is that if a discontinuity is crossed twice within an integrator time step, or if two separate discontinuities lie close together, they may be missed.  Park and Barton address this shortcoming by presenting an approach to search the entire timestep by leveraging interval arithmetic techniques to guarantee detection of all discontinuities~\cite{park1996state}.

Otter et al. discuss the implementation of this zero-crossing function technique in Modelica compilers, where a conservative approach is taken: every relation (with operators such as less than, greater than etc.) between continuous variables is assumed to introduce a discontinuity, and therefore automatically triggers a state event unless the modeler mentions otherwise~\cite{otter1999hybrid}. Fredriksson et al. present an implementation in Assimulo that can be attached to any integrator~\cite{fredriksson2014discontinuities} and is compatible with the FMI standard. In addition, rather than relying on a sign-change, they adopt a domain-based formulation whereby a change between the domains $g < 0$ and $g \geq 0$ is detected. This approach handles a failure mode of the sign-change test for the case where a discontinuity is triggered by the function reaching zero and staying there (as can happen in practical examples such as cutting power to a clutch). In other related work, Tummescheit and Eborn implement crossing functions for multiphase thermal-fluid flow with chemical reactions (with external property models) that allow numerically robust detection of new phases by using a "deviation index" which is the normalized distance to a phase-separation surface~\cite{tummescheit2002chemical}. 

In the present work, several fundamental ingredients of the solver algorithm are taken from this existing literature.  We also use zero-crossing functions are used to mark discontinuities which are then detected by relying on a sign-change, and the location of the discontinuity is determined by performing a root finding search on the integrator's dense interpolant. However, the solver algorithm implements several unique features. First, the solver uses the zero-crossing functions to discover a discontinuity even on step rejections, not merely on step acceptances as is done in alternative approaches. This scheme, designed to handle the anticipated high error in rejected steps, is a key contribution of this paper. For example, the retried step size after rejection is set at half the rejected step size, which empirically yields the best results for the cases studied. Finally, the proposed algorithm is fully composable with any adaptive methods (such as \texttt{Rodas5P}, \texttt{Tsit5}, and \texttt{RadauIIA5} in the Julia SciML ecosystem).

We make the following contributions: 
\begin{enumerate}
    \item We implement a discontinuity-aware ODE and DAE solver that uses domain-specific (user-declared) information to compute appropriate time steps for more efficient integration. This solver is provided open-source in the Julia SciML ecosystem within \texttt{DifferentialEquations.jl} \cite{rackauckas2017differentialequations}. Existing schemes only take advantage of such information on accepted steps, while we extend the scheme to function on rejected schemes as well. 
    \item We apply this solver algorithm to both a scalar ODE with two discontinuities as well as a vapor-compression cycle. We show that discontinuity-awareness results in significantly fewer step rejections and improved solver times, particularly with the \texttt{Rodas5P} solver at loose tolerances, compared to solvers using standard error control methods.
    \item Lastly, we demonstrate how the domain-specific discontinuity information can be specified symbolically in a declarative modeling language (such as Dyad) and passed on to a numerical solver ecosystem.
\end{enumerate}

The paper is organized as follows: Section 2 presents the two exemplar systems, consisting of a scalar ODE system with two discontinuities and a vapor-compression cycle model with varying compressor controls after a zero-flow startup.  Section 3 describes the discontinuity-aware solver algorithm and contrasts with the standard solver, Section 4 presents the results on the two case studies, and Section 5 presents conclusions and directions for future work.

\section{Exemplar Systems}

\subsection{Scalar ODE system}
First, we study the following discontinuous ODE system with two state discontinuities: 
\begin{equation}
    \dot{x} = f(x, t) = \begin{cases}
        2x,        & x < 1 \\
        x + 0.2,   & 1 \leq x < 2 \\
        -4x + 12,  & x \geq 2
    \end{cases}
\end{equation}
with initial condition $x(0) = 0.1$ and $t \in [0,\, 2.5]$. 

\begin{figure}[t]
    \centering
    \includegraphics[width=\columnwidth]{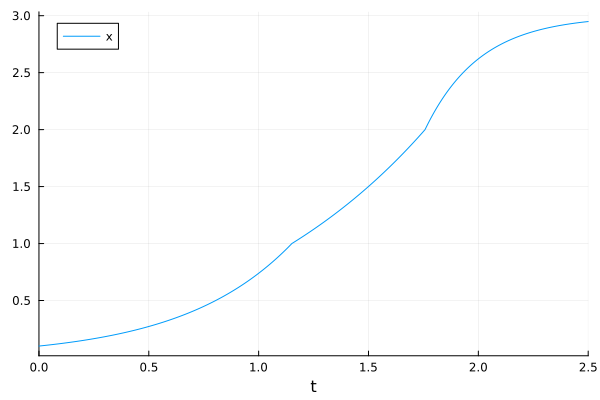}
    \caption{Solution to the toy problem. Note the effects of the derivative discontinuity as the solution crosses $x = 1$ and $x = 2$.}
    \label{fig:toy_problem_sol}
\end{figure}

The solution is plotted in Figure~\ref{fig:toy_problem_sol}, illustrating the fact that the derivative discontinuities produce kinks in the solution at $x = 1$ and $x = 2$. This simple system allows us to explain the stepping dynamics and differences versus a standard ODE solver, as discussed in Section~\ref{result:scalar-ode}.

\subsection{Vapor-compression cycle model} \label{sec:cycle}

\begin{figure*}[t]
    \centering
    \includegraphics[width=0.7\textwidth]{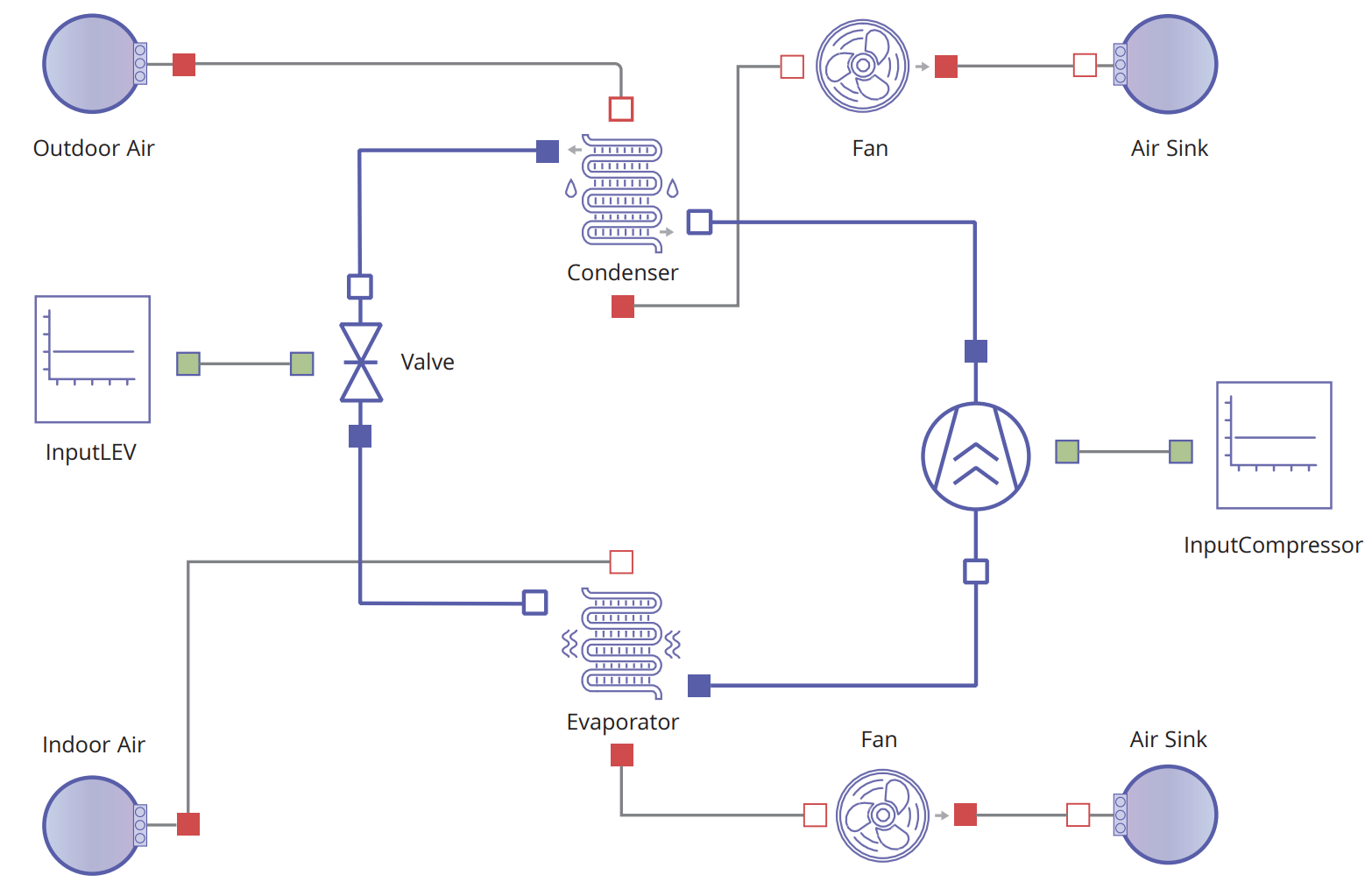}
    \caption{Vapor-compression cycle model in Dyad}
    \label{fig:hvac-dyad}
\end{figure*}

Figure \ref{fig:hvac-dyad} presents a vapor-compression cycle modeled in Dyad. It consists of a positive-displacement compressor, an evaporator and a condenser modeled as a tube-fin heat exchangers with a cross-flow configuration, and a linear expansion valve. The compressor and valve are described via algebraic relations since their dynamics are much faster than the rest of the system. The tube-fin heat exchanger comprises a 1D refrigerant thermal-fluid flow model with a finite-volume discretization along the flow axis consisting of mass, momentum and energy balance equations. Pressure and enthalpy states are selected as differential states to be integrated, while other refrigerant properties are computed from these states using fitted spline-based property models. The heat transfer through the tube walls to the air side is modeled using appropriate efficiency correlations, and the moist air flow has a cross-flow configuration over the finned tubes. A complete description of model is available in~\cite{laughman2017influence} with a summary in the Appendix.  

The Dyad model is developed using the \texttt{Dyad HVAC Components} library \cite{dyad_hvac}, in which reusable subcomponents are developed and interconnected. The Dyad compiler then reduces the full model to the form of a semi-explicit DAE~\eqref{eq:dae} through alias
elimination, index reduction, and tearing. This DAE is then passed to a numerical solver from an  ecosystem such as \texttt{DifferentialEquations.jl} \cite{rackauckas2017differentialequations} to solve the equations of the form 
\begin{align}
\dd{\mathbf{x}}{t} &= \mathbf{f}\bigl(\mathbf{x},\mathbf{y};u(t)\bigr) \\
\qquad
\mathbf{0} &=\; \mathbf{g}\bigl(\mathbf{x},\mathbf{y};u(t)\bigr),
\label{eq:dae}
\end{align}
with $u(t)$ representing the controller inputs (such as compressor speed and valve position), $\mathbf{x}$ representing the differential states (such as refrigerant enthalpies and pressures in each control volume), and $\mathbf{y}$ denoting algebraic states (e.g., heat duties, temperatures and moist-air states). For this case study, compressor speed inputs are chosen so as to drive the system through a cold-start, low-load, peak-load, and partial-shutdown regimes in succession without ever reaching a steady-state, so that several system states repeatedly traverse the refrigerant's two-phase region where discontinuities arise. 

Dyad provides a native mechanism for attaching continuous events to an assembled system model. These events are specified declaratively using Dyad's symbolic language, and thus provide an ergonomic interface for the modeler to communicate domain-specific information to the ODE/DAE solver (\ref{lst:dyad-events}). These continuous events are then passed as callback functions to the integrator and are triggered when the integrated state satisfies the corresponding predicate. Mathematically, these can be written in terms of guard functions $g_j(\mathbf{x})$ which indicate points of discontinuity. We attach two such guards in each refrigerant control volume, one at each phase boundary:
\begin{align}
g_\mathrm{bub}(\mathbf{x}) &= h - h_\mathrm{bub}(P) = 0 \nonumber \\
g_\mathrm{dew}(\mathbf{x}) &= h - h_\mathrm{dew}(P) = 0,
\label{eq:phase-guards}
\end{align}
where $h_\mathrm{bub}(P)$ and $h_\mathrm{dew}(P)$ are the bubble and dew enthalpies evaluated at the particular pressure $P$, and $h$ is the enthalpy state.

\begin{lstlisting}[caption={Dyad syntax for declaring the
phase-boundary events of Eq.~\eqref{eq:phase-guards} in a refrigerant
medium model. Parameters, variables and property relations are omitted.},
label={lst:dyad-events}]
component RefrigerantMedium
  # parameters go here
  # variables go here (including P and h)
relations
  # property relations go here
  discontinuityAt(h = h_bub(P))
  discontinuityAt(h = h_dew(P))
end
\end{lstlisting}

In this case study, we use eight control volumes per heat exchanger giving rise to thirty two events. The discontinuity-aware solver evaluates this guard information if a step is rejected, which is useful in determining subsequent candidate time steps. 
\begin{figure*}[t!]
  \centering
  \includegraphics[width=0.8\textwidth]{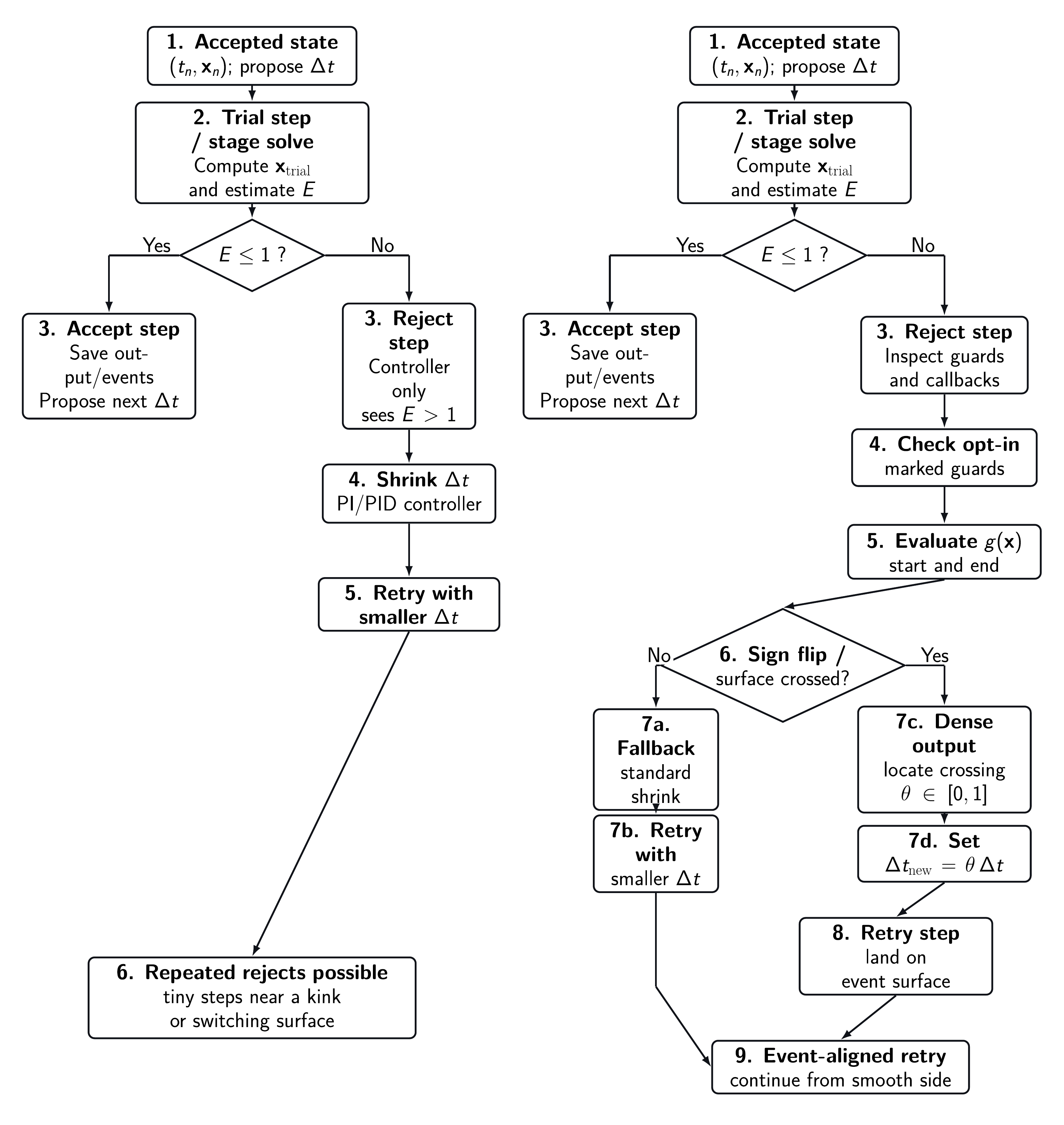}
  \caption{Adaptive time-step control with and without discontinuity-aware logic. The left branch applies the standard rejected-step controller. The right branch uses the marked guard functions together with a nonlinear solve on the dense interpolant to recompute a candidate timestep on the earliest detected discontinuity crossing.}
  \label{fig:discontinuity-flowchart}
\end{figure*}

\section{Discontinuity-Aware Solver \\ Algorithm}\label{sec:method}

The discontinuity-aware solver is partially inspired by the work on delay differential equations (DDEs) by Guglielmi and Hairer~\cite{guglielmi2008}, which proposes a scheme for handling the breaking points in DDEs. These breaking points can be treated similarly to discontinuities in the right hand side of ODEs and DAEs, motivating the scheme presented here. 

The new scheme exposes discontinuity information to the step-size controller thereby enabling cleaner navigation around them. The right flowchart in Figure \ref{fig:discontinuity-flowchart} presents the improved adaptive time-step control logic, which is the core feature of the discontinuity-aware solver.  In comparison, the left flowchart presents the time-step control without discontinuity-awareness.  Of particular note is the fact that the current state of the software only runs on accepted steps, and this is our baseline for comparison. The novelties we introduce are manifest in the behavior on rejected steps, as described below. 

Both solvers first proceed by taking a trial step as appropriate for the method, giving a test value $\hat x$. After taking the step, the solver uses a lower-order Runge-Kutta solution formed from the same stage evaluations to derive a lower-order estimate $\tilde x$. A normalized error estimate for the step is thus derived as
\begin{equation}
    E = \sqrt{\frac{1}{n}\sum_{i=1}^{n} \left(\frac{\hat{x}_i - \tilde{x}_i}{\text{atol} + \text{rtol} \cdot \max(|\hat{x}_i|, |\tilde{x}_i|)}\right)^2}. \vspace{0.1in}
\end{equation}
In the event that the error is less than 1, both schemes will accept the step and proceed to the next step. However, the key difference occurs on step rejections, which we will now further explore.

\subsection{Traditional step-size controllers}
Many step-size controller schemes exist, but the general idea of using heuristic measures to predict the appropriate step size for the next step is often consistent from scheme to scheme. For instance, the Gustafsson controller scheme \cite{gustafsson1991} computes the new step size via the following heuristic:
\begin{equation}
    h_{\text{new}} = \frac{f_{\text{safe}} \cdot h}{E^{1/(p+1)}},
\end{equation}
where $f_\text{safe}$ is a safety factor typically taken around $0.9$, $h$ was the current step size attempted, $E$ was the error of the current step, and $p$ is the order of the method. 

This scheme works well when the system is smooth, but derivative discontinuities lead these heuristic estimates to perform very poorly, with clusters of step rejections around the discontinuities. For this reason, we introduce our discontinuity-aware step-size controller to handle these situations more robustly. 
    
\subsection{Discontinuity-aware step-size controller}

 
We expose discontinuities via callbacks ~\cite{diffeq_callbacks}, which describe discontinuity functions in two main forms:

\begin{enumerate}
     \item Continuous callbacks: A single discontinuity function $g(x)$, which triggers an event at points where $g(x) = 0$.
     \item Vector Continuous callbacks: A vector of conditions, allowing multiple discontinuity surfaces to be simultaneously monitored. 
\end{enumerate}

These are exposed to the step size controller via the integrator interface, allowing the identification of discontinuities in our scheme.  Upon opting to reject a step, we perform the following analysis on the callbacks:

\begin{figure*}[t!]
    \centering
    \begin{subfigure}[t]{0.8\textwidth}
        \centering
        \includegraphics[width=\textwidth]{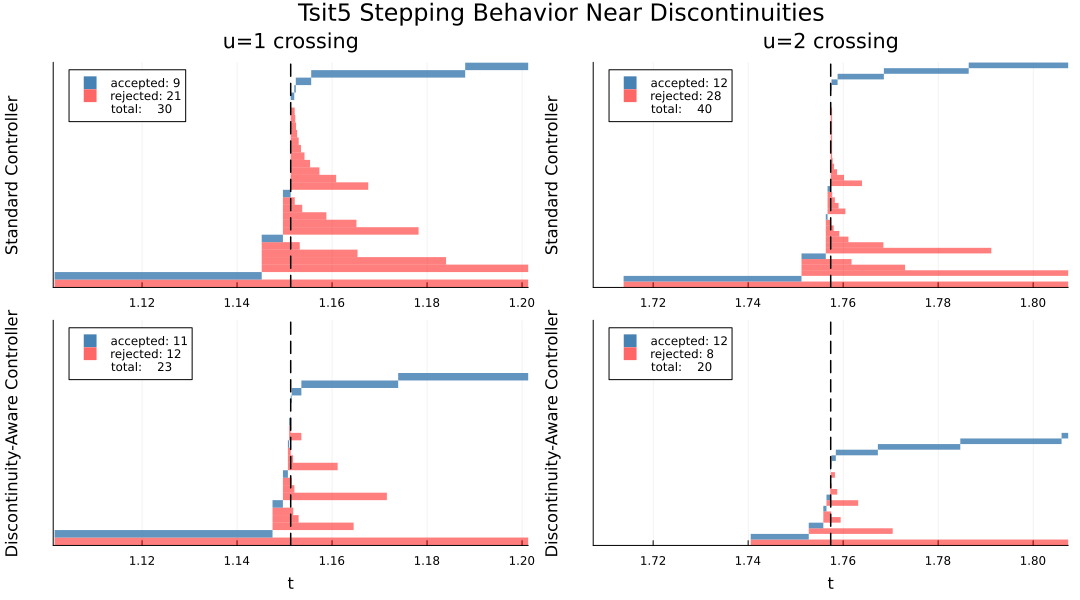}
        \caption{\texttt{Tsit5}}
    \end{subfigure}
    \vspace{0.5em}
    \begin{subfigure}[t]{0.8\textwidth}
        \centering
        \includegraphics[width=\textwidth]{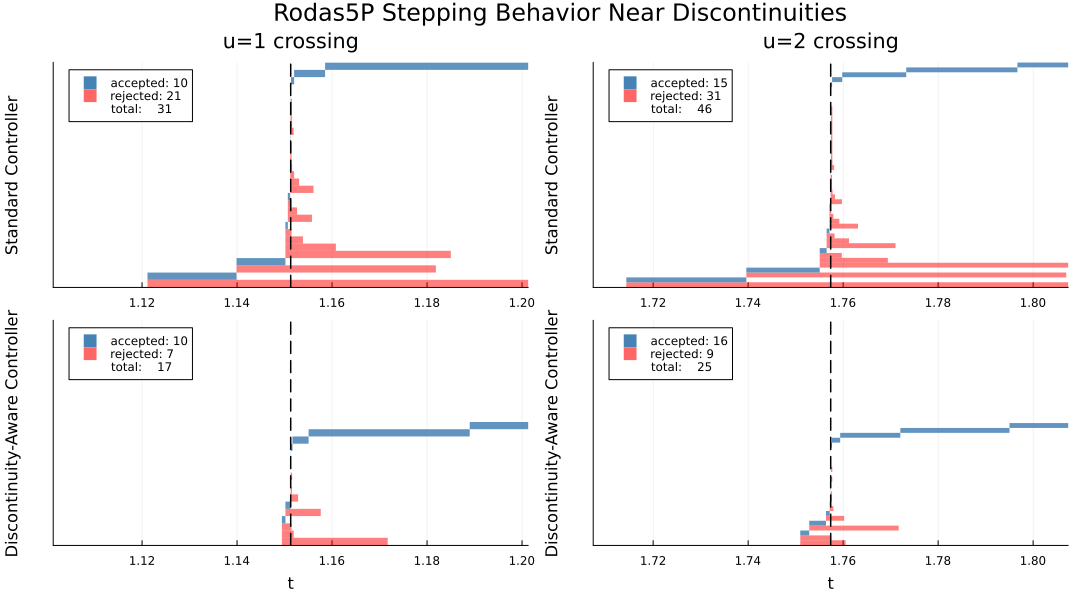}
        \caption{\texttt{Rodas5P}}
    \end{subfigure}
    \caption{Time steps of the \texttt{Tsit5} solver and the \texttt{Rodas5P} solver. The y-axis shows the number of trial time steps, each horizontal bar spans a proposed time step with the bars classified by whether the time step was accepted (blue) or rejected (red). The locations of the discontinuities are marked by the vertical dashed line (note that these are only known in the state space; the corresponding location of the discontinuity in time is not known before solution)}
    \label{fig:stepping}
\end{figure*}
 
\paragraph{Identifying event occurrences:} Consider a step from $t$ to $t + h$, with $f(t) = x_t$, $f(t+h) = x_{t+h}$. For a given callback $g(x)$, a root of $g$ occurs if $g(x_t)$ and $g(x_{t+h})$ have different signs by the intermediate value theorem. Then, $g(x_t) \cdot g(x_{t+h}) < 0$ provides a sufficient condition for detecting the occurrence of an event. We evaluate this guard on all callbacks and only proceed to the subsequent steps on those that had event occurrences; if no events are detected, we fall back to the standard step-size rejection scheme previously described.

\paragraph{Identifying the event locations:} For state-dependent callbacks previously described, we seek the time $t_{\text{event}}$ at which $g(f(t_{\text{event}})) = 0$. This time is determined by using the dense interpolant for the solver, either derived specially for the integrator or using a Hermite interpolation, and an interval nonlinear solve is done to determine the location of the event using a bisection-like search. This gives a point in time at which the event occurred, and we choose the first such event to step to directly. We note that the dense interpolant is built from data already computed during the rejected trial step, so does not require either full function evaluations, Jacobian evaluations or matrix factorizations. However, given that the error was already high on the step, the interpolant will not be perfect, and hence we expect some numerical error with the computation of the event, but even with this error, this scheme causes the timestep to be chosen much closer to the discontinuity than a standard PI timestepper which is unaware of the discontinuity's existance. 

 Given this error, we choose to mildly bias the result of the solve, $\theta$, via the following formula:

 \begin{equation*}
     \theta = (0.5 + 0.4 \cdot \sin(\pi \cdot (\theta - 0.5)))
 \end{equation*}
 
 Or equivalently in polynomial form (to reasonable error, for faster computational purposes), 

 \begin{equation*}
    \theta = 0.5 + 1.2372 \cdot (\theta - 0.5) - 1.7487 \cdot (\theta - 0.5)^3
 \end{equation*}

 The motivation for this biasing is that we want each step to maximize the information yielded, hence an extremely large step (close to the prior rejected step) or a very small step is suboptimal. Therefore, we choose to try and bias our steps towards roughly $\texttt{dt} / 2$, while also maintaining use of the nonlinear solvers results. 
 
\paragraph{Resumption of Smooth Integration:} Upon crossing the discontinuity, we reset error history of the controller. This is done to indicate the resumption of integration in a new domain, in which prior assumptions may not apply. This allows us to more efficiently determine the appropriate step-sizes for this regime. 

In addition to this error reset, we also restrict the first step after a discontinuity forced step. This is done out of caution, given that we expect that we have just navigated the discontinuity, but in the event that we have not, we want a limited step to again maximize the information gain from this rejected step. 

More technically, suppose our first rejected step ends at time $t$. Then, after our discontinuity step, we restrict the subsequent step to $\min(dt, (t - t_{curr}) / 4$, where $dt$ is the proposed step-size. Thus, if our intended step takes us beyond the area where we think the discontinuity is, to prevent unnecessary failures, we limit this first step before continuing integration as normal. 

We further note that this discontinuity-aware time-step controller is composable with a variety of solvers in the \texttt{DifferentialEquations.jl} \cite{rackauckas2017differentialequations} ecosystem.

\section{Results and Discussion}

In this work, we study the application of this scheme to three algorithms: \texttt{Rodas5P} (a Rosenbrock method applicable to both DAEs and ODEs), \texttt{RadauIIA5} (a fully-implicit Runge-Kutta method applicable to both DAEs and ODEs), and \texttt{Tsit5} (an explicit Runge-Kutta method applicable to ODEs) for both of the exemplar models.

We note that in both the standard and discontinuity-aware solver configurations, the solver has access to zero-crossing functions at all times. The distinction is that the standard controller only uses this information on accepted steps, while our scheme additionally exploits it on rejected steps.

\subsection{Scalar ODE system}
\label{result:scalar-ode}

Figure \ref{fig:stepping} presents the time step plots of the \texttt{Tsit5} and \texttt{Rodas5P} solvers with a discontinuity-aware controller versus a standard controller. On each plot, the left pane shows the time steps taken near the $u = 1$ discontinuity while the right pane shows this near the $u=2$ discontinuity. The y-axis shows the number of trial time steps. Each horizontal bar spans a proposed time step with the bars classified by whether the time step was accepted (blue) or rejected (red). Studying the right pane of Figure \ref{fig:stepping} (a), we observe that the first step is rejected in both cases. Both the discontinuity-aware and standard solvers shrink the next proposed time step. However, only the discontinuity-aware solver computes an appropriately-sized time step; while it is still rejected, gives this controller much more information to proceed. The standard solver shrinks the time step proportional to the error of the step, resulting in the rejection of the next time step being rejected before the succeeding one is accepted. This pattern repeats several times as we note that the standard solver has significantly more (red) rejected time steps. This trend also continues with Figure \ref{fig:stepping} (b) as we observe that the discontinuity-aware solver requires significantly fewer trials to pinpoint the discontinuity and integrate past it so as to solve the ODE efficiently. 


\begin{figure*}[t!]
    \centering
    \makebox[1\textwidth][c]{%
        \begin{subfigure}[t]{0.58\textwidth}
            \centering
            \includegraphics[width=\textwidth]{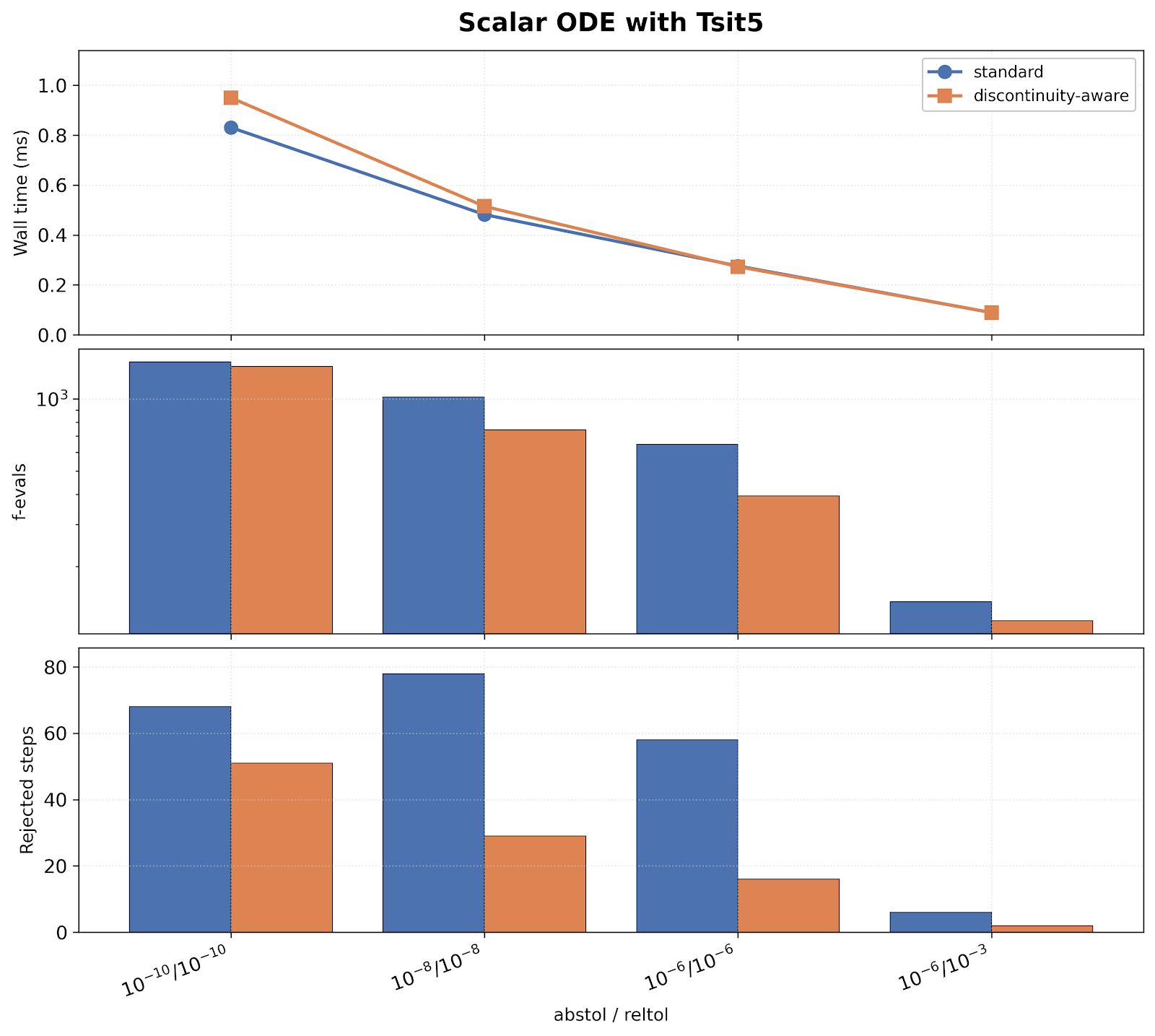 }
        \end{subfigure}%
        \hfill 
        \begin{subfigure}[t]{0.58\textwidth}
            \centering
            \includegraphics[width=\textwidth]{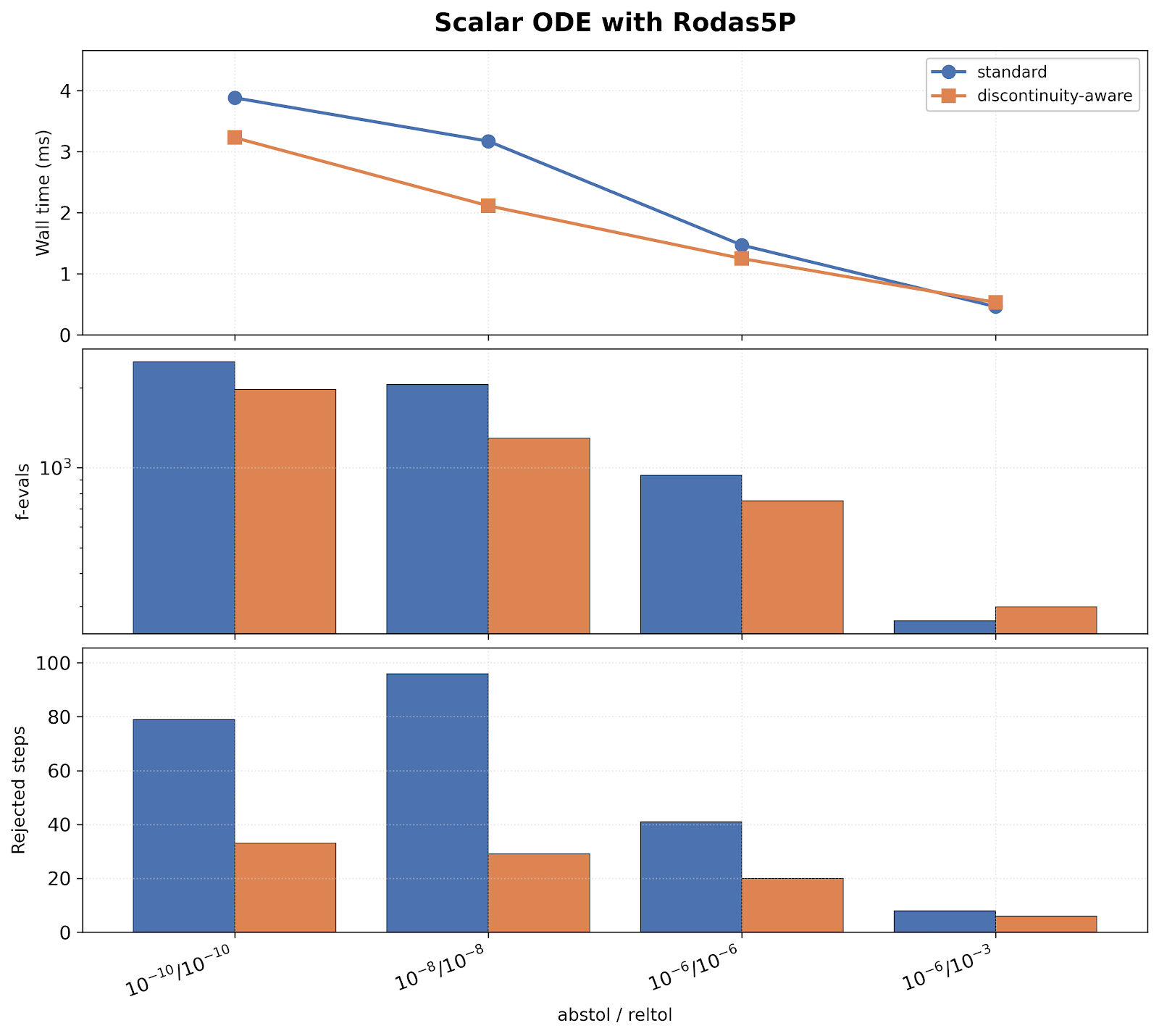}
        \end{subfigure}%
    }
    \caption{Wall time (ms), number of function evaluations and number of rejected steps at different absolute and relative tolerances with a discontinuity-aware and standard time-step controller. Metrics for \texttt{Tsit5} are shown on the left while \texttt{Rodas5P} are on the right.}
    \label{fig:scalar-results}
\end{figure*}

\begin{figure*}[t!]
     \centering
     \begin{subfigure}[t]{0.48\textwidth}
         \includegraphics[width=\textwidth]{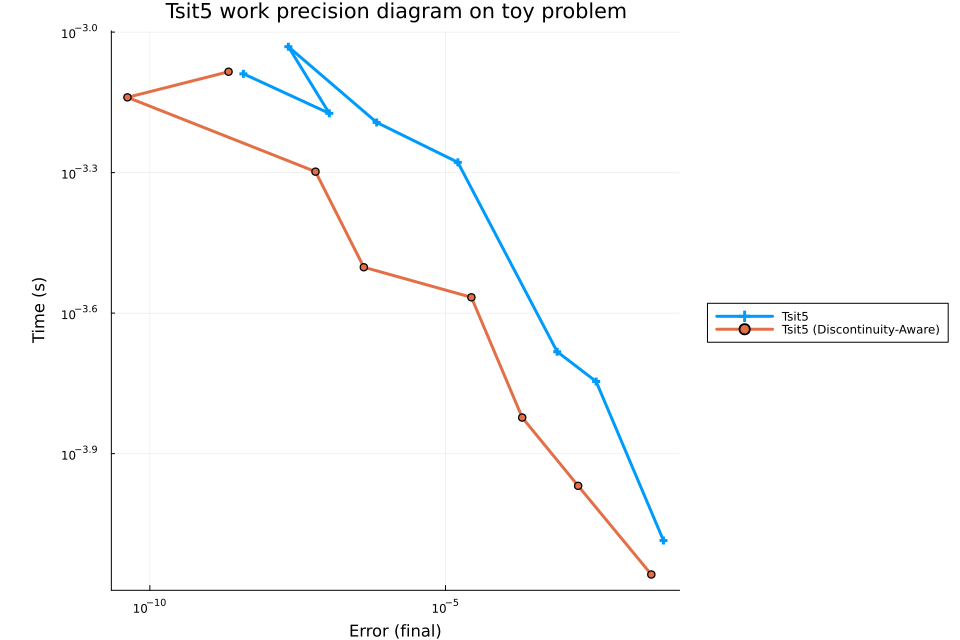}
         \caption{\texttt{Tsit5} work-precision diagram on the toy problem.}
     \end{subfigure}
     \hfill
     \begin{subfigure}[t]{0.48\textwidth}
         \includegraphics[width=\textwidth]{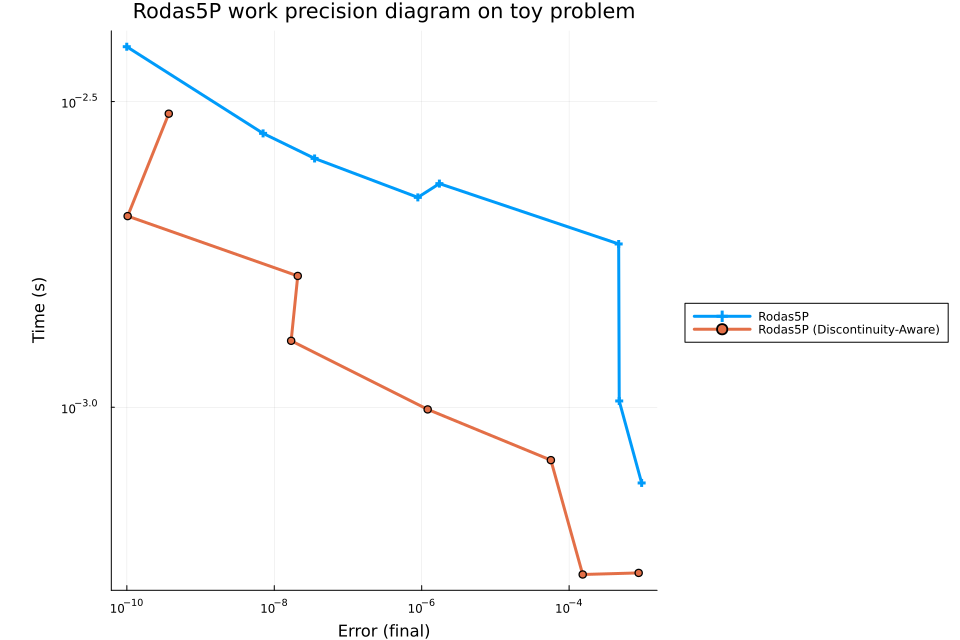}
         \caption{\texttt{Rodas5P} work-precision diagram on the toy problem.}
     \end{subfigure}
     \caption{Work-precision diagrams for \texttt{Tsit5} and \texttt{Rodas5P} on the toy problem. Both plots are taken with $\text{reltol}$ ranging from $10^{-2}$ to $10^{-9}$, and $\text{abstol}$ ranging from $10^{-4}$ to $10^{-11}$. The vertical axis shows solve time and the horizontal axis shows the final error in the solution.}
     \label{fig:toy-wp}
 \end{figure*}

Figure \ref{fig:scalar-results} presents several solver metrics: wall time (ms), number of function evaluations, and number of rejected steps for both \texttt{Tsit5} and \texttt{Rodas5P} solvers with  discontinuity-aware and standard time-step controllers. For both \texttt{Tsit5} and \texttt{Rodas5P}, we note that discontinuity-awareness results in significantly fewer rejected steps. Similarly, discontinuity-awareness results in fewer function evaluations which are a major contributor to solver times. Figure~\ref{fig:toy-wp} presents a work-precision diagram: The x-axis denotes the L2-norm error of the solution versus a ground-truth solution while the y-axis denotes the solver time. The results on the work-precision diagrams build on this as the performance of the discontinuity-aware solver is consistently faster than the standard solver.  

An interesting point to note is that the benefit for the implicit method, \texttt{Rodas5P}, appears larger than the benefit for the explicit method, \texttt{Tsit5}. This gives useful insight into when the discontinuity detection scheme adds value: on an implicit method, with a higher cost per rejected step, the discontinuity-aware step-size controller has a positive effect. On the other hand, while the scheme still helps for an explicit solver, we see a more modest improvement, as the rootfinding overhead introduced by the discontinuity detection scheme is non-trivial relative to the actual step cost.

\subsection{Vapor-compression cycle}

\begin{figure*}[t!]
    \centering
    \makebox[1\textwidth][c]{%
        \begin{subfigure}[t]{0.58\textwidth}
            \centering
            \includegraphics[width=\textwidth]{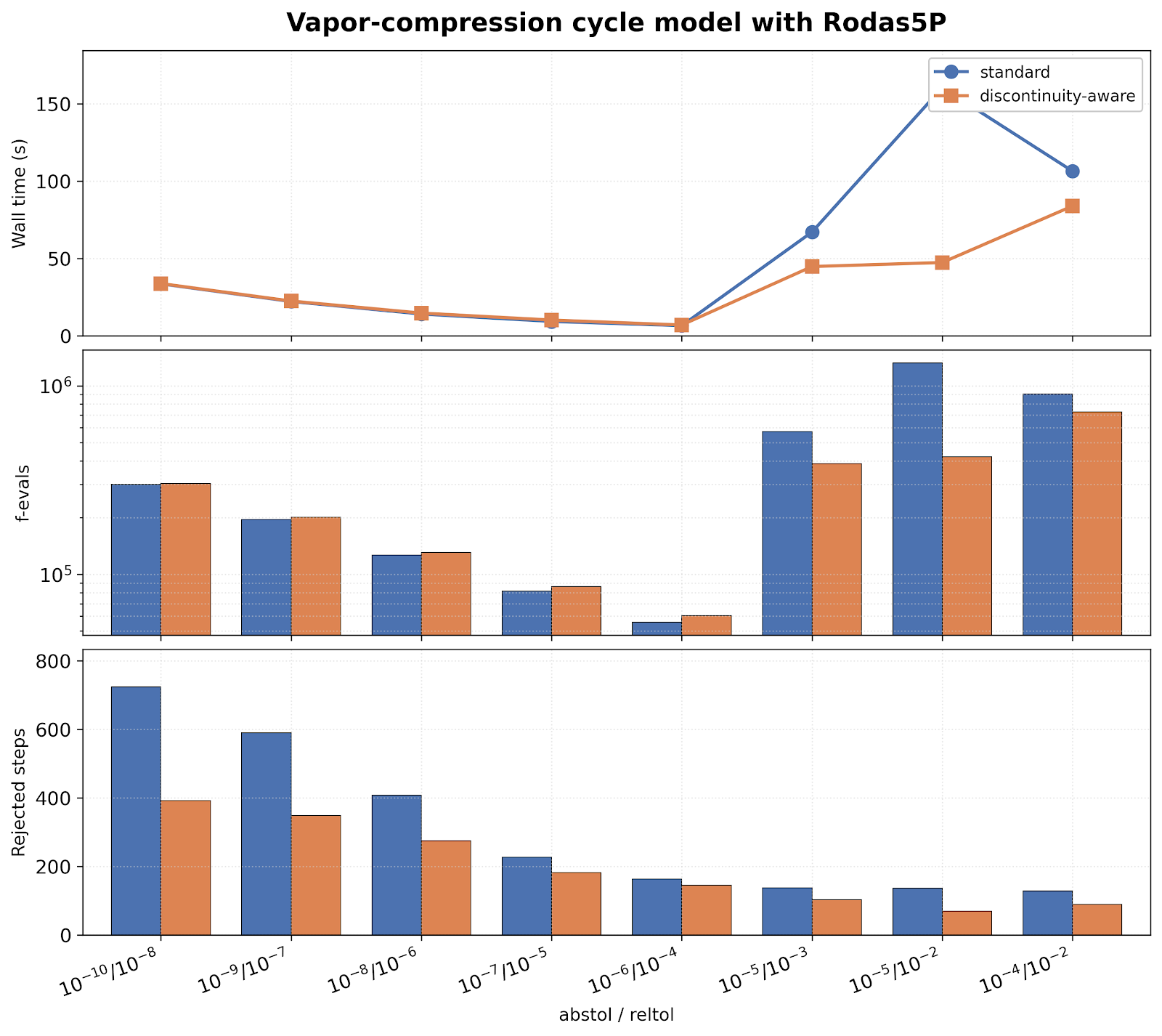}
        \end{subfigure}%
        \hfill 
        \begin{subfigure}[t]{0.58\textwidth}
            \centering
            \includegraphics[width=\textwidth]{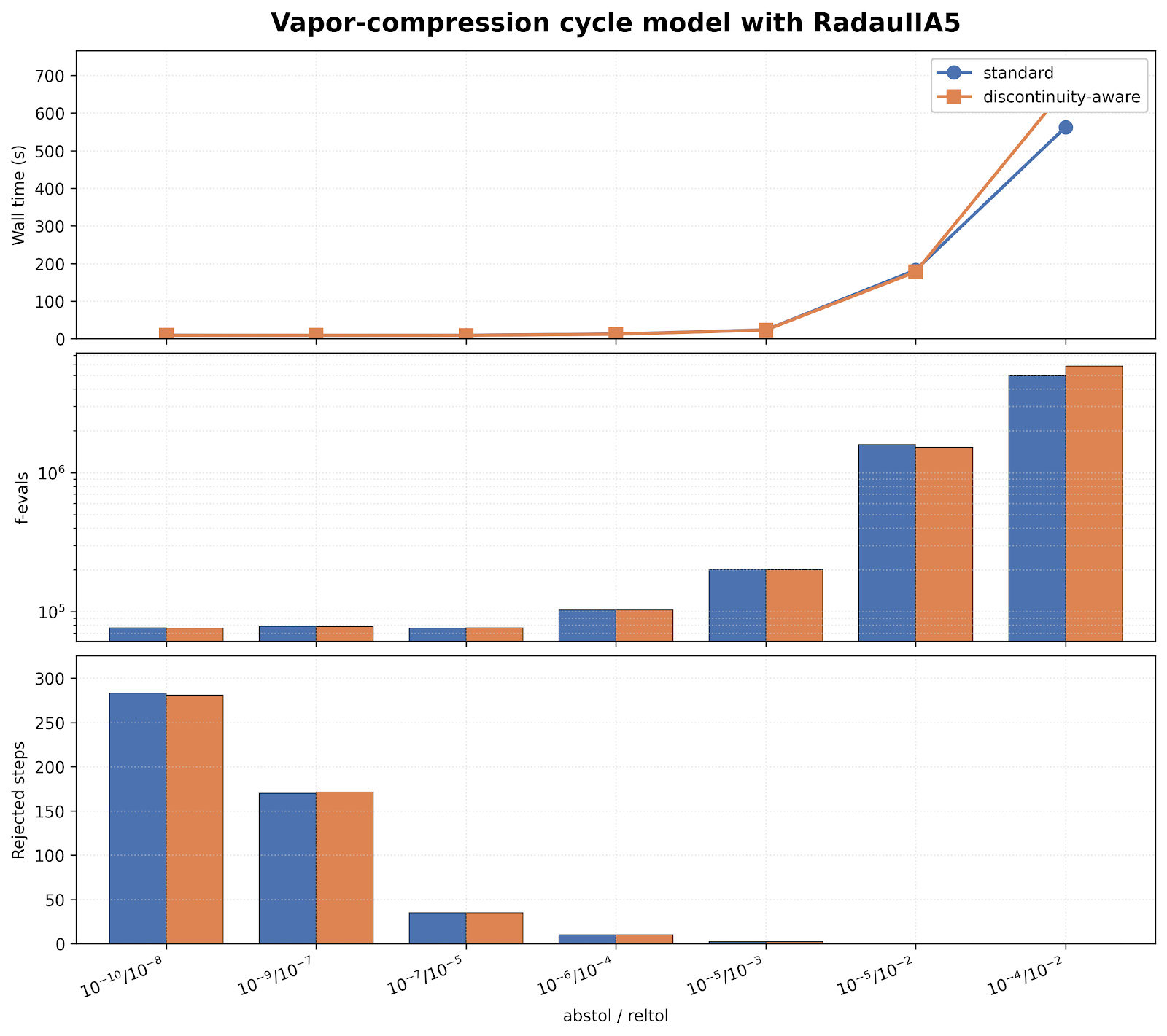}
        \end{subfigure}%
    }
    \caption{Wall time (ms), number of function evaluations and number of rejected steps at different absolute and relative tolerances with a discontinuity-aware and standard time-step controller. Metrics for \texttt{Rodas5P} are shown on the left while \texttt{RadauIIA5} are on the right. Metrics are calculated as the average over 15 runs with varying initialization conditions.}
    \label{fig:hvac-results}
\end{figure*}

Figure~\ref{fig:hvac-results} presents the solver metrics for the vapor-compression cycle with the \texttt{Rodas5P} and the \texttt{RadauIIA5} solver. Note that our results are shown as averages over 15 runs at each tolerance level, as the solver performance is strongly dependent on the solver initialization; some solves perform poorly with discontinuity-awareness and cleanly otherwise, while the opposite also occurs. We therefore analyze general trends for the clarity of results, rather than the individual run variance. 

We first analyze the results for \texttt{Rodas5P}. Considering the loose tolerance regimes, we observe performance improvements with the discontinuity-aware step-size controller, with 2x speedups on average across the low tolerances. The poor performance of the standard solver in these regimes likely stems from the fact that the standard controller takes large steps at loose tolerances and incurs high error around discontinuities.  While the loose tolerances allow the solver to accept steps despite high error near discontinuities, the large error estimates cause the controller to dramatically shrink subsequent steps (even though the steps are accepted), leading to inefficient navigation of the discontinuity region.

Further consideration of these results suggests that the full value of the discontinuity-aware scheme lies in its ability to prevent "stagnations", or larger clusters of tiny accepted steps around discontinuities, which cause solve times to blow up. This behavior was seen in the toy problem above, and the table below shows the tail behavior of these test runs on the full model. 

\begin{table}[ht]
\centering
\resizebox{\columnwidth}{!}{%
\begin{tabular}{|l|rr|rr|rr|rr|}
\hline
 & \multicolumn{2}{c|}{\textbf{Median}} & \multicolumn{2}{c|}{\textbf{P75}} & \multicolumn{2}{c|}{\textbf{P90}} & \multicolumn{2}{c|}{\textbf{Max}} \\
\textbf{abstol / reltol} & \textbf{ND} & \textbf{D} & \textbf{ND} & \textbf{D} & \textbf{ND} & \textbf{D} & \textbf{ND} & \textbf{D} \\
\hline
Default                 & 13.2 &  9.8 &  38.8 &  13.3 & 445.8 &  81.5 &  916.9 & 804.5 \\
$10^{-4}$ / $10^{-2}$  & 42.0 & 69.1 & 132.2 & 117.1 & 168.5 & 170.4 &  651.3 & 217.3 \\
$10^{-5}$ / $10^{-2}$  & 34.7 & 32.5 & 122.9 &  50.0 & 322.7 &  98.8 & 1245.9 & 150.5 \\
$10^{-5}$ / $10^{-3}$  & 22.6 & 24.9 &  59.8 &  27.8 & 223.0 &  30.5 &  304.3 & 380.4 \\
$10^{-6}$ / $10^{-4}$  &  6.3 &  6.9 &   6.4 &   7.0 &   6.9 &   7.2 &    9.4 &   7.6 \\
\hline
\end{tabular}%
}
\caption{Rodas5P solve time (s) distributions over 15 runs on the complete HVAC model. ND = standard controller, D = discontinuity-aware controller. P75 represents the 75th percentile, while P90 is the 90th percentile.}
\label{tab:rodas5p-timing-stats}
\end{table}

Strong evidence of the discontinuity-aware scheme's benefits at high tolerances is manifest in this tail behavior. For example, the default tolerance is 5x faster at the 90th percentile of runs, and $10^{-5}/10^{-3}$ is 8x faster at the same percentile. Furthermore, the maximum solve time at $10^{-5}/10^{-2}$ is likewise 8x faster with discontinuity-awareness. The 75th percentile is about 2x faster across all of the tolerances. 

While the differences are negligible in the median performance, but the discontinuity-aware scheme performs significantly better when failures and difficulties occur. This serves as key evidence for the benefits of this scheme, as the bounds for the worst case solve time are tighter than they are for the existing schemes. The successful recovery of a solve that took 20 minutes at $10^{-5}/10^{-2}$ in only 2.5 minutes with our scheme is the strongest indicator of this claim.

In comparison, we see that the discontinuity-aware scheme still rejects far fewer steps with \texttt{Rodas5P} at tight tolerances, but to negligible benefit, and even minor slowdowns. At these tight tolerances, both schemes are forced to take smaller steps; this setting favors the standard scheme, as the cost of rejections decreases. The benefits of our algorithm are minor here, but the overhead is also limited.

We note an interesting contrast between the results with \texttt{Rodas5P} and \texttt{RadauIIA5}. The differences between the standard and discontinuity-aware solvers are negligible across the board with \texttt{RadauIIA5}, with solve times being close to identical and step rejection percentages likewise very similar. Interestingly, the discontinuity-aware scheme often increases both step acceptances and step rejections, but the cost of this appears to be mitigated with each individual step being cheaper. This is apparent when considering the function evaluations with \texttt{RadauIIA5}, as there are minimal increases and even decreases in function evaluations despite the increase in steps. This is likely because of the more efficient navigation of discontinuities by the discontinuity-aware scheme, allowing the Newton iterations to converge more easily and hence reducing function evaluations. 

At tight tolerances, \texttt{RadauIIA5} appears to be more robust than \texttt{Rodas5P}. This difference can be understood as a product of the fact that \texttt{RadauIIA5} is a fully implicit Runge-Kutta method, while \texttt{Rodas5P} is linearly implicit. \texttt{Rodas5P} benefits dramatically from discontinuity-awareness at loose tolerances because its single linearization per step is particularly vulnerable to discontinuities, as it takes large steps and hence incurs severe error. On the other hand, \texttt{RadauIIA5} shows little benefit because the full Newton iteration more effectively handles the discontinuity internally, converging to a valid solution even across discontinuities, at the cost of more iterations but without catastrophic step rejection. Again, this is likely because both schemes are already exposed to discontinuities through the callback mechanism, and differ in how they utilize this information on rejected steps. For fully implicit methods like \texttt{RadauIIA5}, this appears to yield minimal benefit, while there are large gains for linearly implicit methods like \texttt{Rodas5P}. At tight tolerances, both take smaller steps, yielding similar results.

\section{Conclusions}

We have presented a discontinuity-aware ODE and DAE solver in which user-declared information is used to compute appropriate time steps at discontinuities, and is now available in the open-source \texttt{DifferentialEquations.jl} ecosystem. We also demonstrated how this information can be specified symbolically in the Dyad declarative modeling language and passed to the solver.

Benchmarking on a scalar ODE and a vapor-compression cycle DAE model shows that the scheme yields significantly fewer step rejections and improved solver times, with the largest gains observed for \texttt{Rodas5P} at loose tolerances, where we see speedups of up to 8x in the tail behavior of HVAC models. The contrast between \texttt{Rodas5P} and \texttt{RadauIIA5} reveals an important practical guideline: discontinuity-awareness provides the greatest benefit for linearly implicit methods at loose tolerances, where large step sizes make the solver particularly vulnerable to derivative discontinuities, while fully implicit methods are more naturally robust. Furthermore, the analysis of \texttt{Rodas5P} shows the critical value of the discontinuity-aware scheme in preventing large solve failures, which  somewhat mitigates the challenges of difficult integration problems.



\printbibliography

\appendix
\section{Vapor-compression cycle model}
\label{app:cycle}
\subsection{Tube-fin heat exchangers}

Both the condenser and the evaporator are tube-fin heat exchangers comprised of the refrigerant-side flow, the tube wall, and the air-side flow.

\paragraph{Refrigerant flow.}
The refrigerant flow is 1D thermal-fluid flow. Mass, momentum, and energy conservation are then governed by
\begin{align}
\pd{(\rho A)}{t} + \pd{(\rho A v)}{x}   &= 0, \label{eq:ref-mass}\\
\pd{(\rho v A)}{t} + \pd{(\rho v^2 A)}{x} &= -A\,\pd{P}{x} - F_f, \label{eq:ref-mom}\\
\pd{(\rho u A)}{t} + \pd{(\rho v h A)}{x} &= vA\,\pd{P}{x} + vF_f + \pd{Q}{x}.
\label{eq:ref-eng}
\end{align}
Discretising each refrigerant circuit yields an
ordinary differential equation of the form for the mass of volume
$j$, for instance,
\begin{equation}
\dd{(\rho_j V_j)}{t} \;=\; \mdot_k - \mdot_{k+1},
\label{eq:ref-mass-cell}
\end{equation}
with the finite volume enthalpy flow evaluated by upwind descretization. The frictional pressure drop in both the
single- and two-phase regions is taken as a scaled quadratic relation,
\begin{equation}
\Delta P \;=\; K\,\frac{(\Delta P)_0}{\mdot_0^{\,2}}\,\mdot^{\,2},
\label{eq:ref-friction}
\end{equation}
and the refrigerant heat-transfer coefficient is assigned a constant value
within each phase. The derivative discontinuities at the bubble and dew
curves discussed in Section~\ref{sec:cycle} enter the assembled
DAE~\eqref{eq:dae} through refrigerant property computations.

\paragraph{Tube wall.}
Conductive heat transfer through the property wall takes the form 
\begin{equation}
\begin{aligned}
(m_w c_w + m_f c_f)\frac{\mathrm{d}T_w}{\mathrm{d}t}
&= \alpha_\mathrm{ref} A\,(T_r - T_w) \\
&\quad + \dot{m}_a\!\bigl[c_{p,a}(T_{a,\mathrm{in}} - T_{a,\mathrm{out}}) + \\
&(\omega_{a,\mathrm{in}} - \omega_{a,\mathrm{out}})\,\Delta h_{fg}\bigr].
\end{aligned}
\label{eq:wall}
\end{equation}

\paragraph{Air-side flow.}
The air stream crosses the fin side and exchanges both sensible
and latent heat with the wall using replaceable heat and mass transfer correlations

\subsection{Compressor}

The compressor is a positive-displacement machine modelled algebraically,
with no mass or energy storage. Its behaviour is set by two performance
maps fitted to manufacturer data --- a volumetric efficiency, which fixes
the refrigerant mass flow, and an isentropic efficiency, which fixes the
discharge enthalpy:
\begin{align}
\eta_v(f, p_r, \ldots) &= \frac{\mdot_\mathrm{comp}}{\rho_\mathrm{suc}\, V_f}, \label{eq:comp-vol}\\
\eta_\mathrm{is}(f, \ldots) &= \frac{h_\mathrm{dis,isen} - h_\mathrm{suc}}
                                     {h_\mathrm{dis} - h_\mathrm{suc}},
\label{eq:comp-isen}
\end{align}
where $V_f$ is the swept volume rate at shaft frequency $f$,
$h_\mathrm{dis,isen}$ is the discharge enthalpy of the equivalent
isentropic compression, $\rho_\mathrm{suc}$ is the suction-side density,
and both efficiencies are maps in $f$ and the pressure ratio $p_r$.

\subsection{Expansion valve}

The LEV is modeled as an isenthalpic, algebraic orifice without mass or
energy storage. The mass flow follows the standard orifice relation
\begin{equation}
\mdot \;=\; C_v(\theta)\,A_v\,\sqrt{\rho_\mathrm{in}\,\Delta P},
\label{eq:valve}
\end{equation}
in which $C_v(\theta) A_v$ is the effective flow coefficient --- a smooth,
differentiable function of the valve position $\theta$ ---
$\rho_\mathrm{in}$ is the upstream density, and $\Delta P$ is the pressure drop across the valve.







\end{document}